\documentclass[10pt,conference]{IEEEtran}
\IEEEoverridecommandlockouts
\makeatletter
\newcommand{\newlineauthors}{%
  \end{@IEEEauthorhalign}\hfill\mbox{}\par
  \mbox{}\hfill\begin{@IEEEauthorhalign}
}
\makeatother

\usepackage{cite}
\usepackage{amsmath,amssymb,amsfonts}
\usepackage{algorithmic}
\usepackage{graphicx}
\usepackage{textcomp}
\usepackage{xcolor}

\usepackage{url}
\usepackage{multirow}

\usepackage{soul}

\usepackage[most]{tcolorbox}

\usepackage{colortbl}

\usepackage{float}

\def\BibTeX{{\rm B\kern-.05em{\sc i\kern-.025em b}\kern-.08em
    T\kern-.1667em\lower.7ex\hbox{E}\kern-.125emX}}
\begin{document}

\title{The Role of the Retrospective Meetings in Detecting, Refactoring and Monitoring Community Smells\\{\LARGE A Qualitative Study}}

%\author{\IEEEauthorblockN{Anonymous Authors}}
\author{

\IEEEauthorblockN{Carlos Dantas}
\IEEEauthorblockA{%\textit{dept. name of organization (of Aff.)} \\
\textit{Federal Institute of Rio Grande do Norte}\\
Apodi, Brazil \\
carlos.dantas@ifrn.edu.br}

\and
\IEEEauthorblockN{Tiago Massoni}
\IEEEauthorblockA{%\textit{dept. name of organization (of Aff.)} \\
\textit{Federal University of Campina Grande}\\
Campina Grande, Brazil\\
massoni@computacao.ufcg.edu.br}

\and

\IEEEauthorblockN{Camila Sarmento}
\IEEEauthorblockA{%\textit{dept. name of organization (of Aff.)} \\
\textit{Federal University of Campina Grande}\\
Campina Grande, Brazil\\
camilasarmento@copin.ufcg.edu.br}

\newlineauthors

\IEEEauthorblockN{Rayana Rocha}
\IEEEauthorblockA{%\textit{dept. name of organization (of Aff.)} \\
\textit{Federal University of Campina Grande}\\
Campina Grande, Brazil\\
rayana@copin.ufcg.edu.br}

\and
\IEEEauthorblockN{Danielly Gualberto}
\IEEEauthorblockA{%\textit{dept. name of organization (of Aff.)} \\
\textit{Federal University of Campina Grande}\\
Campina Grande, Brazil\\
daniellygualberto@copin.ufcg.edu.br}
}

\maketitle

\begin{abstract}
\textit{Background.} Retrospective meetings play a vital role in agile development by facilitating team reflection on past work to enhance effectiveness. These meetings address various social aspects, including team dynamics, individual performance, processes, and technologies, ultimately leading to actions for improvement. Despite their importance, limited research has explored how these meetings handle forms of social debt, particularly Community Smells --- recurring dysfunctional patterns in team dynamics, such as poor communication or isolated work practices. \textit{Goal.} This study seeks to understand how retrospective meetings address a few core Community Smells, examining whether these meetings help identify smells, make it possible to formulate refactoring strategies, support the monitoring of refactoring actions, and contribute to preventing the most prominent Community Smells. \textit{Method.} We conducted semi-structured interviews with 15 practitioners from diverse organizations who regularly participate in retrospective meetings. The interviewees shared their experiences with retrospectives, the challenges discussed, and subsequent improvement actions. The study focused on the four most cited Community Smells in the literature --- Lone Wolf, Organizational Silo, Radio Silence, and Black Cloud. Data was analyzed iteratively using a priori coding to examine Community Smells and inductive open coding inspired by Grounded Theory. \textit{Findings.} The findings indicate that retrospective meetings indeed enable the identification of core Community Smells. However, while strategies for refactoring are often formulated, their implementation and monitoring remain inconsistent. Additionally, an emphasis on positive aspects during these meetings may help in preventing Community Smells. \textit{Conclusion.} This study offers valuable insights to practitioners and researchers, highlighting the importance of addressing social debt in software development within agile practices.
\end{abstract}


\section{Introduction}

%context - comm. smells
In Software Engineering, the term \textit{Social-Debt} describes unforeseen project costs, which are related to the non-ideal conditions of the software development team \cite{tamburri2013}.
One of the main sources of social debt is \textit{Community Smells}, which refers to social and organizational circumstances with implicit causal relationships to potential problems in the software development process \cite{tamburri2013}. 
Examples include \textit{Organizational Silo} (OS), where communication difficulties and challenges in sharing project knowledge arise between subgroups within the team, and \textit{Lone Wolf} (LW), characterized by team members working in isolation with little or no communication with others. Other recurring examples of Community Smells are: \textit{Black Cloud} (BC) -- when organizations do not provide the conditions for social interactions and effective communication between teammates, not supporting the exchange of knowledge as professional experience or understanding of projects in progress; and \textit{Radio Silence} (RS) (also known as \textit{Bottleneck}), a scenario where team communication structures are not conducive to spread information across the teams efficiently -- for example, a person working as a unique information intermediary, leading to communication overload and massive delays~\cite{catolino2020, tamburri2015, tamburri2021, palomba2021b}. More than 30 Community Smells have been cataloged in other studies, highlighting a diverse set of socio-technical challenges that impact software development processes and team dynamics~\cite{caballero2023}.

%context - ret meetings
Community Smells can be identified in various ways and at different stages, additional to occur in the different stages of the software development process. In agile teams, Retrospective meetings are considered an important practice in agile methods \cite{agilepraticeguide2017, agiletrenches2015, stateagile2016} whose purpose is for the team to reflect on the work done in the last iteration and adjust to become more efficient in the next iteration \cite{scrumguide2020}.
In the literature, there are several suggestions on how to conduct retrospective meetings \cite{kua2012retrohandbook, derby2006agileretrospectives, caroli2020funretrospectives}. Among the points discussed in the retrospective meeting, aspects related to the project, technologies, process, team, and individuals stand out, making the retrospective meeting the most comprehensive and focused on continuous improvement by the team compared to the other meetings in the agile Scrum methodology \cite{martini2019, agiletrenches2015, agilepraticeguide2017, scrumguide2022bok}.
Given the focus on reflection and improvement, retrospective meetings present a potential opportunity for teams to identify and address Community Smells more than other meetings in Scrum, such as Dailies and Reviews.

%problem - gap in the literature
Studies on Community Smells have addressed causes and effects, as well as the identification of new Community Smells and their detection \cite{Tamburri2019a, tamburri2021}. However, few studies focus on strategies for dealing with such circumstances or use retrospective meetings to investigate social debt in agile teams.

%solution
This exploratory study aims to examine the role of retrospective meetings in detecting, refactoring, monitoring, and preventing Community Smells. 
The study focuses specifically on four core Community Smells commonly cited in the literature: Lone Wolf, Organizational Silo, Radio Silence, and Black Cloud.
%method
The methodology employed in the study is a qualitative study based on semi-structured interviews with 15 practitioners who have experience with retrospective meetings or similar ceremonies for continuous improvement. Participants were asked about their experiences with the four core Community Smells using vignettes based on established descriptions. Data analysis was iterative and involved a combination of a priori coding to examine Community Smells and inductive open coding inspired by Grounded Theory~\cite{charmaz2009}.  

%main findings
The study found that retrospective meetings can help identify Community Smells in agile development teams. 
This detection was triggered by the negative impacts of Community Smells on individual or team performance, prompting discussions and the need for improvement.
Also, results show that teams frequently formulate strategies for refactoring these Community Smells during the retrospective meetings, although implementation and monitoring of these strategies remain inconsistent. 
Despite the challenges in consistently addressing Community Smells, the study also found that retrospective meetings could contribute to their prevention.

%consequences and implications
The findings have a number of consequences for both researchers and practitioners in the field of software engineering.
For practitioners, it emphasizes the importance of not only identifying Community Smells during retrospective meetings but also developing concrete action plans to address them and monitoring the effectiveness of these actions. For researchers, we point out challenges and hypotheses to be studied in more detail.

\section{Related Work}
\label{relatedwork}

Among the refactoring actions indicated in the literature are team restructuring, defining and/or restructuring communication plans, and mentoring team members \cite{catolino2020, sarmento2022, tahsin2022}. However, despite the presence of these refactoring strategies, several challenges remain. For instance, Caballero-Espinosa, Carver, and Stowers~\cite{caballero2023} highlighted in a systematic literature review the existence of Community Smells without indication of refactoring strategies, strategies with unproven or partially proven effectiveness, and a lack of details about the strategies indicated in the literature.

This study sought to identify strategies defined in the retrospective meetings for refactoring Community Smells. However, due to the limitation of the Community Smells addressed, it does not present details of the execution or effectiveness of the strategies, results that would contribute to the gaps identified by Caballero-Espinosa, Carver, and Stowers\cite{caballero2023}.
Other studies reveal the effects of agile software development practices on social factors within teams, highlighting, for example, that retrospective meetings are associated with work engagement and trust among team members \cite{law2005, mchugh2012, muller2021}. Frequent and open communication within the team, knowledge sharing, and obtaining feedback are factors that impact trust among members and are related to the practice of retrospective meetings \cite{mchugh2012}.

The importance of retrospectives as a tool for continuous improvement is evident in agile development contexts. A longitudinal study examined retrospective meeting reports from teams in Large-Scale Agile Development to identify problems and improvement actions, aiming to understand how retrospectives are conducted and how improvement actions could be enhanced\cite{dingsoyr2018}. The research identified 109 points, of which 65 were areas where the team could make improvements. However, while many of these points were categorized under ``People and Relationships'', this category received limited attention in the final analysis because the authors evaluated theses as less related to how the large-scale agile development methodology influences teams.

Similarly, Martini, Stray, and Moe~\cite{martini2019} conducted a study to characterize the issues identified in retrospective meetings of large-scale agile development teams. The case study consisted of interviews, observations, and document analysis to investigate what teams discuss in intra-team and inter-team retrospective meetings concerning technical debt, social debt, and process debt. The reported results show that most of the problems in intra-team retrospectives are related to social debt, which suggests the need for further research.
In this context, retrospective meetings are highlighted as a data source for investigating different aspects of software development. Unlike such works \cite{martini2019, law2005, mchugh2012, muller2021, dingsoyr2018}, this study has a specific focus on the practice of retrospective meetings and their role in refactoring social debt, particularly in the circumstances known as Community Smells, showing how these smells are highlighted, what strategies are defined and which monitoring and prevention are carried out through the retrospective meetings.

\section{Methodology}
\label{method}

\subsection{Study Objective and Research Questions}

%objective
For this research, a qualitative study was proposed to explore how Community Smells manifest in retrospective meetings and how improvement actions impact these smells. 
Constructivism was adopted as the epistemological stance, recognizing that knowledge is constructed through social interactions and individuals' interpretations of their experiences \cite{creswell2014}.

The study addresses the following research questions:

\textit{RQ1: Do Retrospective Meetings help identify Community Smells?}

Although socio-technical issues are highlighted in the retrospective meetings, it is not known how specific situations cataloged as Community Smells are highlighted and whether they can be identified in these meetings.

\textit{RQ2: Are Strategies to Refactoring Community Smells defined in Retrospective Meetings?}

Focusing on continuous improvement, teams define action plans to be executed in the following iterations. In this sense, the objective is to investigate whether the actions defined in the retrospective meetings include strategies to refactor Community Smells.

\textit{RQ3: Do Retrospective Meetings Enable the Monitoring of Actions for Refactoring Community Smells?}

When starting a retrospective meeting, some teams review the highlights of their past retrospective meeting, evaluating defined action plans as a form of monitoring. Therefore, it is desired to investigate if the teams include the strategies defined for refactoring Community Smells in this monitoring and how it occurs.

\textit{RQ4: What are the benefits of Retrospective Meetings in preventing Community Smells?}

The importance of retrospective meetings for agile teams is recognized, given the focus on continuous improvement and possible discussion of different points. From this perspective, it is possible to identify the benefits of the retrospective meeting regarding the prevention of Community Smells, primarily considering defined actions and the highlight of positive aspects by the team.

\subsection{Data collection}

In this study, the semi-structured interview model was adopted, where the researcher is not limited to the initially defined questions and can use additional questions to obtain more information on the desired topics \cite[p. 425]{sampieri2013}.

The interview protocol\footnote{Supplementary Material: \url{https://dx.doi.org/10.6084/m9.figshare.26371006}} was designed to explore the participants' experiences in retrospective meetings concerning the highlighted issues, the improvement actions defined, and the positive aspects indicated by the teams in these meetings. 
The interview protocol, among other questions, includes: \textit{``How retrospective meetings work (sprint retrospective) of your team?''}, \textit{``What problems are reported in the retrospective meetings of your team?''} and \textit{``What improvement actions have been defined?''}.

For this study, we focused on Lone Wolf (LW), Organizational Silo (OS), Radio Silence (RS) and Black Cloud (BC)\cite{tamburri2013}, as they were reported more often in past studies \cite{sarmento2022, catolino2020, tahsin2022}, presenting more consolidated definitions on causes, effects and refactoring practices among the Community Smells so far cataloged. 
 
The focus on a reduced number of Community Smells considered the length of interviews and the impact on participant engagement.

Participants were asked about the four Community Smells situations using vignettes based on descriptions by Caballero-Espinosa, Carver, and Stowers~\cite{caballero2023}. 
The use of vignettes, which are brief, descriptive scenarios, to present situations of Community Smells offers several advantages over simply naming the smell. Vignettes provide a concrete context for participants to understand and identify the specific behaviours and dynamics associated with each smell. Directly naming a Community Smell like ``Lone Wolf'' might be too abstract or potentially embarrassing for participants, especially if they recognize themselves or their colleagues in the description.
For example, a vignette used for Lone Wolf:

\begin{quotation}
    \textit{``In any retrospective, has it ever been mentioned if any team member works more in isolation, with little or no communication with the other members?''}
\end{quotation}

The first three interviews were conducted as a pilot study, during which no modifications to the prepared protocol were identified.
All interviews were online (via Google Meet\footnote{\url{https://meet.google.com}}), recorded and transcribed using the FireFlies platform\footnote{\url{https://fireflies.ai}}, with a subsequent manual review of the transcriptions using the OTranscribe tool\footnote{\url{https://otranscribe.com}}.

\subsection{Recruitment}

This study used semi-structured interviews with practitioners who, in their respective roles, have participated in retrospective meetings or ceremonies with similar purposes (continuous improvement).
Initial participant recruitment (first iteration collection-analysis) was done by convenience sampling. Subsequently, theoretical sampling \cite[p.139]{charmaz2009} was employed, seeking new participants who could contribute data for the research topic's development and theoretical refinement, considering gender, professional experience and role/function in the team for the diversity of participants.
Recruitment of new team members in different roles, with diverse professional backgrounds and genders, was conducted by searching, analyzing profiles, connecting with candidates, and exchanging messages on the LinkedIn\footnote{\url{https://linkedin.com}} platform.
The research was approved by the Research Ethics Committee %of the Alcides Carneiro University Hospital (HUAC)
, and all participants were contacted and informed about the research objectives to give their free consent for the interview, in accordance with the guidelines of the National Health Council (CNS).% of Brazil.

\subsection{Analysis}

%MASSONI: talvez seja melhor usar uma figura para ilustrar esses ciclos
The analysis process was carried out iteratively, interspersed with the data collection phase across four iterations, to achieve the theoretical sampling. Thus, each iteration of collection-analysis was decided by certain profiles of new participants, as described below:

\begin{enumerate}
\item There were three participants, all developers. After the pilot interviews, it was decided to recruit participants with more experience time and different roles.

\item In this round, four practitioners participated. They were quality analysts and developers with different professional experience times; however, all were men. For the next round, we decided to diversify gender and get a vision of the leadership of teams, in addition to recruiting continuously different roles and experience time.

\item Four practitioners were recruited, half of them being leaders and women in different roles. For the next round, we focus on balancing gender and leadership functions.

\item Finally, different roles were sought in the team and again by leaders and women. The four new participants did not bring new perceptions or codes during coding.
\end{enumerate}

\begin{figure*}
    \centering
    \includegraphics[width=1.0\linewidth]{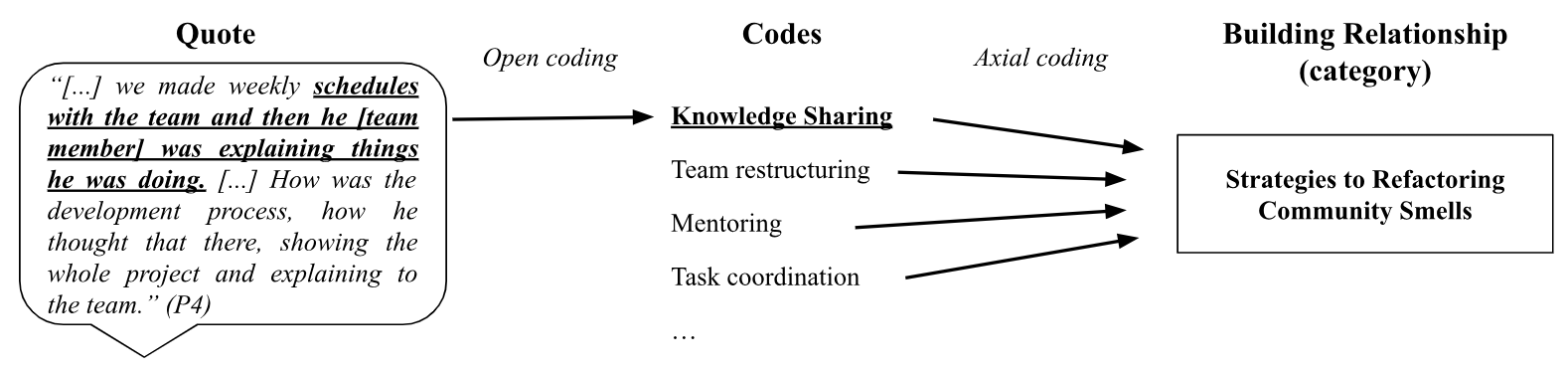}
    \caption{The coding process applied to the analysis to answer the research questions}
    \label{fig:coding}
\end{figure*}

The analysis consisted of a priori coding to explore situations of the Community Smells (related to RQ1) and inductive open coding and axial coding inspired by the Grounded Theory\cite{charmaz2009, creswell2014} to answer the other research questions (RQ2--4). A Fig.~\ref{fig:coding} shows an example of the coding process, in which segments of the interviews were coded and, subsequently, the codes were categorized into themes related to the research questions by using axial coding -- that aims to classify, synthesize and organize large amounts of data and regroup them in new ways after open coding \cite{creswell2014}. Fig. 1 shows one specific quote derived from the question \textit{``What improvement actions have been defined?''} for community smells approached, in this case, the strategy defined for Black Cloud.

MAXQDA\footnote{\url{https://www.maxqda.com}} was used for the coding process, constant comparison, and categorization of the identified themes related to the research questions.

Coding was carried out by the first researcher, and the segments of the interviews related to the occurrence of Community Smells were also coded by the third researcher as a form of redundancy if the situations reported by the interviewees characterized one of the Community Smells. The second researcher was involved in the coding process to resolve disagreements.

\section{Results}
\label{sec:resultados}

Fifteen practitioners from different agile software development teams working in various organizations were interviewed. Table~\ref{tab:perfis} presents the distribution of the participants by gender, role, leader function (team leader, tech lead, quality lead) and years of professional experience in software development teams, split by interview-analysis cycle.

\begin{table}[ht]
\caption{Profile of the Interviewees}
\begin{center}
\begin{tabular}{|c|c|c|c|c|} \hline
    \textbf{ID} &
    \textbf{Gender} &
    \textbf{Role} & 
    \textbf{Leader} &
    \textbf{Experience (Y)} \\ \hline %\hline
    
    P1 & Man & Developer & No & $<2$ \\ %\hline
    P2 & Man & Developer & No & 5 -- 10 \\ %\hline
    P3 & Woman & Developer & No & 2 -- $<$5 \\ \hline %\hline \hline
    P4 & Man & Developer & No & 2 -- $<$5 \\ %\hline
    P5 & Man & Quality Analyst & No & 5 -- 10 \\ %\hline
    P6 & Man & Quality Analyst & No & 2 -- $<$5 \\ %\hline
    P7 & Man & Developer & No & 5 -- 10 \\ \hline %\hline \hline
    P8 & Woman & Quality Analyst & Yes & 2 -- $<$5 \\ %\hline
    P9 & Man & Quality Analyst & No & 5 -- 10 \\ %\hline
    P10 & Man & Product Owner & No & $<$2 \\ %\hline
    P11 & Woman & Developer & Yes & $>$10\\ \hline %\hline \hline
    P12 & Man & Quality Analyst & No & 5 -- 10 \\ %\hline
    P13 & Man & Developer & No & 5 -- 10 \\ %\hline
    P14 & Woman & Developer & Yes & 5 -- 10 \\ %\hline
    P15 & Woman & Developer & Yes & 5 -- 10 \\ \hline
\end{tabular}
\end{center}
\label{tab:perfis}
\end{table}

According to the participants, in their teams, the retrospective meeting adopts a standard structure in which members indicate \textit{``what was good in the sprint that ended''} and \textit{``what could be improved for the next sprint''}, followed by a discussion and definition of improvement actions (also called action points) to be executed in the following sprints. Only two participants reported that in some meetings, the team used some different dynamics but that the results as to the highlighted points were the same -- positive and negative points occurred in the sprint closed, action points to be performed in the following sprints.

In the following sections, the detailed results of the research questions defined in this study are presented.

\subsection{(RQ1) Do Retrospective Meetings Help Identify Community Smells?}
\label{sec:results_for_A}

Participants were asked, using vignettes, whether situations characterizing the Community Smells known as \textit{Lone Wolf}, \textit{Organizational Silo}, \textit{Radio Silence}, and \textit{Black Cloud} were highlighted in any of their teams' retrospective meetings. At this point, all participants affirmed the presence of at least one of the Community Smells and 7 of 15 participants related two or more. Table~\ref{tab:ocorrencia} presents the distribution of the reports of the occurrence of Community Smells from the conducted interviews.

% Tabela de Ocorrência de Smells -- Entrevistas X Smells
\begin{table}[ht]
    \centering
    \caption{Reports of Community Smells in the Interviews}
    %\resizebox{\textwidth}{!}{
    \begin{tabular}{|c|c|c|c|c|}
        \hline
        \textbf{ID} &
        \textbf{LW} &
        \textbf{OS} & 
        \textbf{RS} &
        \textbf{BC} \\ \hline \hline

        P1 & X & X & & \\ \hline
        P2 & & & X & \\ \hline
        P3 & & X & & \\ \hline 
        P4 & X & X & & X \\ \hline
        P5 & X & & X & \\ \hline
        P6 & & X & & \\ \hline
        P7 & & & & X \\ \hline 
        P8 & X & & X & \\ \hline
        P9 & & & X & X\\ \hline
        P10 & X & & X & \\ \hline
        P11 & X & X & X & \\ \hline 
        P12 & X & X & X & X\\ \hline
        P13 & & & & X\\ \hline
        P14 & & & & X\\ \hline
        P15 & & X &  & \\ \hline
        
    \end{tabular}%}
    \label{tab:ocorrencia}
\end{table}

\begin{figure*}[ht]
    \centering
    \includegraphics[width=0.8\linewidth]{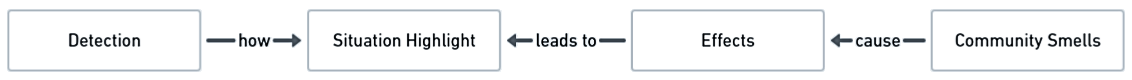}
    \caption{Detection of Community Smells in Retrospective Meetings}
    \label{fig:deteccao-csmells}
\end{figure*}

In general, those smells are detected in retrospective meetings when the situation is highlighted as a problem or bottleneck by at least one of the team members, following the pattern depicted in Fig.~\ref{fig:deteccao-csmells}. 
The emphasis on these situations is driven by the individual perceptions of the members due to the impacts of the effects caused by the Community Smells on individual and/or team work. The sections below detail context and methods of smell detection in the retrospective meetings.

\subsubsection{\textit{Organizational Silo}}

Reports of \textit{Organizational Silo} in retrospective meetings indicate issues such as task overload with task reallocation, rework, delays in task delivery, difficulties integrating parts of features, and perceptions of focus on different projects or products among team members. Participant P4 stated that the situation characterized as \textit{Organizational Silo} was discussed in the retrospective meeting and highlighted the consensus among members about the team size and the fact that it was divided between different demands and focuses. P4 explains that \textcolor{violet}{\textit{``There have been situations where the team became [...] too large. [...] part of the team was focused [...] on one delivery, [...] the team had different scopes within the same team.''}}

Participant P6 reported the Community Smell in question, associating its occurrence with the fact that the team has two main products, in which developers and quality analysts are constantly divided, leading to a focus that causes an implicit division of the team between the products. P6 reports: \textcolor{violet}{\textit{
``This happens a lot [...] because, in fact, we are working on two products. And so, some development members and some QAs end up being more focused on one product.''}}

The cases reveal that task coordination is the main cause of \textit{Organizational Silo}, with effects such as knowledge concentration, focus on different products or projects, and reduced communication and collaboration among members, which led the situation to be reported in the Retrospective Meeting, exemplifying concretely the pattern illustrated in Figure~\ref{fig:deteccao-csmells}.

\subsubsection{\textit{Lone Wolf}}

Situations that characterize \textit{Lone Wolf}, being related to a particular individual, are often described in a more generic manner to avoid embarrassment and shame. Such situations are brought up in the retrospective meeting through the effects caused during the Sprint, such as incidents with features in production, delays in task completion, or delays in feature delivery.
Participant P4 reported a case discussed in the retrospective meeting where one of the team members acted in isolation from the others, making modifications that led to system failures. Participant P4 said: \textcolor{violet}{
\textit{``The person [...] made the modification and did not inform the team, [...] caused a problem where everyone was [...] lost and looking for what had happened. [...] did not properly track the result.''}}

In the interview with P10 (\textit{Product Owner}), who, in some of their team’s retrospective meetings, also acts as a facilitator, they expressed difficulties with communication among some team members, which characterized certain isolation and delayed the handling of development issues, leading to delays and even non-delivery of features defined for the \textit{sprint}. Participant P10 described their perception and the report of the highlighted situation in the team’s retrospective meeting: \textcolor{violet}{
\textit{``One person and another, I perceive that [...] they are more shy or don’t communicate much [...]. Feedback from other team members [...] that we had problems throughout the sprint, [...] the person did not keep the team updated. [...] Some people [...] cannot communicate [...] when there is a problem or escalate it and keep the team updated.''}}

In addition to being cautious when addressing these issues in the retrospective, some interviewees reported that these situations are discussed in one-on-one meetings with their leaders or project managers.

\subsubsection{\textit{Radio Silence}}
Reports of \textit{Radio Silence} in the Retrospective include delays in deliveries, overload on more senior members, and difficulties in development due to dependence on data or access procedures.

Participant P8, as a testing leader, reported feeling overwhelmed due to constant requests from team members, saying: \textcolor{violet}{
\textit{``He is the technical lead of the development team, and he experiences the same situation [...] people have doubts about a story [...] they go to him first. [...] if he can't resolve it [...] they seek out the PO. [...] They lack the proactivity [...] and wait for someone else to solve it.''}}
In the interview with P10, the dependency on other client professionals to obtain/access data is highlighted. P10 reports: \textcolor{violet}{
\textit{``We have many dependencies on the client. So, sometimes, we can't advance [...] because we need access to data. [...] access depends on someone obtaining necessary access or finding the data, which often the client doesn't know where it is.''}}

Participant P6 highlighted difficulties in communicating with the client to obtain the necessary information for development. P6 reported: \textcolor{violet}{
\textit{``We had some requirements, but they were not very well documented [...] we tried to communicate. They asked us to exchange emails, but the information still wasn’t clear. We even had to undo the implementation because [...] was not exactly what the client wanted.''}}
\textit{Radio Silence} situations often involve individuals external to the team, which complicates resolution and makes these issues more persistent throughout the development iterations.

\subsubsection{\textit{Black Cloud}}

It is noted that reports of \textit{Black Cloud} are closely associated with the overload of senior members or the team’s perception of the need to share knowledge among members. Participant P4 reports: \textcolor{violet}{
\textit{``It mainly happened [...] when new people were joining the team, [...] I was one of the people with the most seniority and knowledge [...] then a lot of tasks were coming to me, [...] we realized it would be interesting to spread the knowledge more.''}}
In the interview with P7, the participant reports the team's perception, saying: \textcolor{violet}{\textit{``[...] was mentioned that unit tests needed improvement, and so it was up to one person to plan and manage training sessions about unit tests to guide and clarify doubts.''}}
In general, participants reported having moments of knowledge sharing and openness among members, which facilitate the exchange of experiences and knowledge transfer.

\subsection{(RQ2) Are Strategies to Refactoring Community Smells defined in Retrospective Meetings?}

\begin{figure*}
    \centering
    \includegraphics[width=0.8\linewidth]{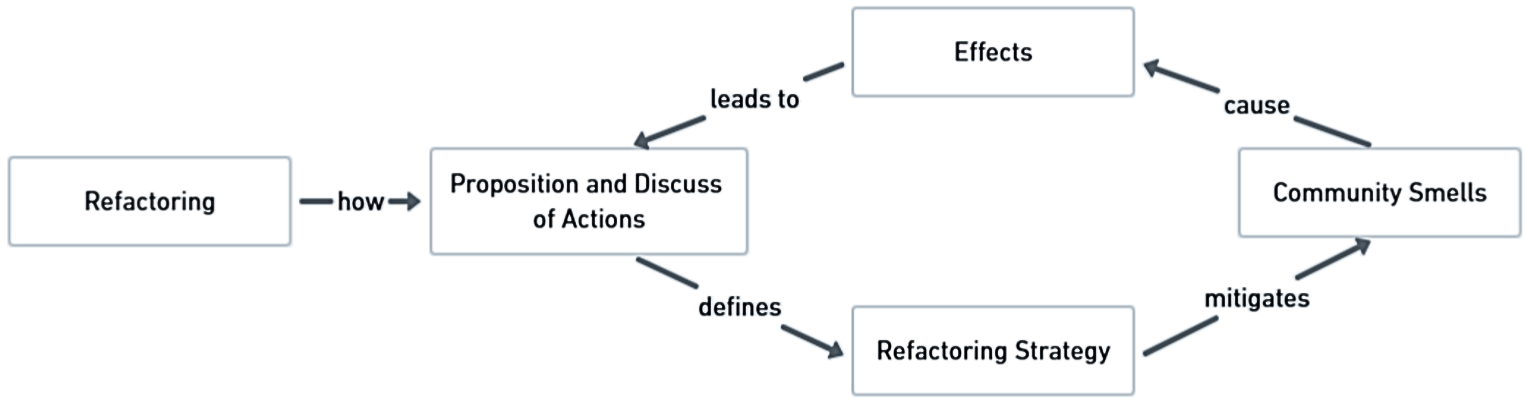}
    \caption{Refactoring of Community Smells in Retrospective Meetings}
    \label{fig:refactoring-csmells}
\end{figure*}

Refactoring is understood as the strategies (actions) defined to be implemented during the following sprint. 
As illustrated in Fig.~\ref{fig:refactoring-csmells}, the definition of strategies generally occurs through the proposition and discussion of actions by members during the retrospective meeting. A concrete example of the definition of strategies for refactoring Community Smells is the report from participant P7 presented at the end of Section~\ref{sec:results_for_A}, in which the team mentions and defines the strategy for improvement in unit tests.

The research participants were asked about which actions were defined to address the Community Smells situations. Table~\ref{tab:mitigacao} presents the strategies identified by the participants, among them, strategies present in the literature, such as team restructuring, the definition of communication plan, and mentoring of team members.

\begin{table*}[ht]
    \centering
    \caption{Strategies for Refactoring Community Smells}
    %\resizebox{\textwidth}{!}{
    \begin{tabular}{|c|l|}
    \hline
        \textit{\textbf{Community Smell}} & \textbf{Refactoring Strategies} Number of participants who cited.)\\ \hline
        \hline
        Organizational Silo &
        \begin{tabular}{l}
             Team restructuring (division) (2)\\

             Contracts between members of specific areas (2)\\
            
            Coordination of team tasks (1)\\

            Improvement in technical and business documentation (1)\\
        \end{tabular} \\ \hline
         Lone Wolf &
          \begin{tabular}{l}
            Mentoring of a member or team on the practice to be followed (3) \\
            Coordination of tasks by team leadership (1) \\
            Improvement in the communication process (board and automatic alerts) (1) \\
        \end{tabular} \\ \hline
         Radio Silence &
          \begin{tabular}{l}
          Definition of communication plan (3)\\
          
          Exposure to other members, teams, and stakeholders (2) % (planilhas de contatos, canais de comunicação, reuniões específicas e incentivos/orientação).
          \\
        Request superiors to intervene in the situation (1)\\
        Mentoring of team members (1)\\
        \end{tabular} \\ \hline
         Black Cloud &
          \begin{tabular}{l}
            Moments of sharing knowledge and experiences among members (6) \\
            Improvement in task coordination (1) \\
        \end{tabular} \\ \hline
    \end{tabular}%}
    \label{tab:mitigacao}
\end{table*}

In general, refactoring strategy is defined (as action plan), however, some situations characterized as \textit{Lone Wolf} are brought up in the Retrospective, but no action plan is defined, remaining merely as a notice to the team, as reported by participant P10: \textcolor{violet}{\textit{
``Some things, indeed, we establish an action plan. Who is responsible and what is to be done. For this specific point, it remained more as a message, just information.''}}

Participant P8 reported that, despite the situation being addressed in the Retrospective, members argued that it was a personality profile issue — \textcolor{violet}{\textit{``In this case, we ended up somewhat defending the person because, for example, they have an introverted profile.''}} However, the participant, being one of the leaders, started mentoring the member to mitigate the situation.
Similarly, participant P5 highlighted the situation characterizing \textit{Lone Wolf} in the Retrospective, but no action was established by the team. Participant P5 said: \textcolor{violet}{\textit{``No, because I think it was more about myself, [...] has something to do with my personality. [...] So, no, we didn't put any action in place.''}}

In cases of \textit{Radio Silence}, it was found that some strategies are associated with stakeholders and other external teams. As a result, some situations remain unresolved, as participant P9 mentioned regarding dependency on third parties for data access: \textcolor{violet}{\textit{``There wasn’t really anything, right? It was simply ``just tell me the CPF [Individual Taxpayer Registry] and I’ll clean the database for you'' and that was it.''}}
Participant P5 also reported that the team requested managers to mediate situations with an external \textit{Product Owner}, from whom the team identified a constant dependency to clarify doubts. This situation led to various delays and rework until the \textit{Product Owner} was replaced. Participant P5 said: \textcolor{violet}{\textit{``It was always the same. Basically, the manager would take it to him (PO), show the retro board where the team, in general, was feeling the lack of a PO, but this never changed. So, it was kind of like there was no answer.''}}

The main strategies identified for \textit{Black Cloud} and \textit{Organizational Silo} situations are present in the literature, such as defining times for knowledge sharing or restructuring the team \cite{tamburri2015, catolino2020, tahsin2022}.

\subsection{(RQ3) Do Retrospective Meetings Enable the Monitoring of Actions for Refactoring Community Smells}

\begin{figure*}
    \centering
    \includegraphics[width=0.85\linewidth]{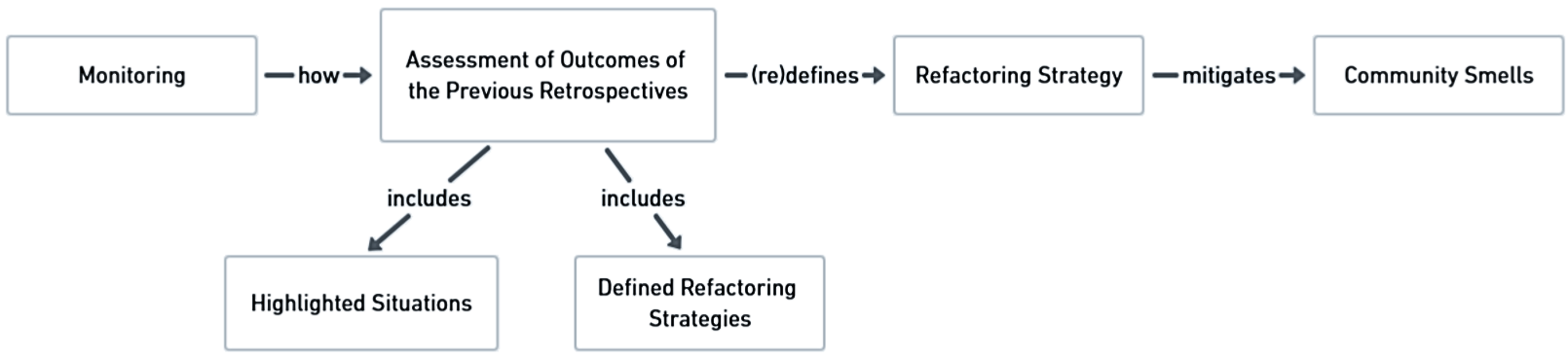}
    \caption{Monitoring of Community Smells in Retrospective Meetings}
    \label{fig:monitoring-csmells}
\end{figure*}

Monitoring of Community Smells in retrospectives, illustrated in Fig.~\ref{fig:monitoring-csmells}, also happens due to members' perceptions of situations in the ending \textit{sprint}, which may be recurring from previous \textit{sprints} and discussed in past retrospectives. Monitoring, therefore, can involve revisiting points from the previous retrospective to assess the situations and actions regarding their effectiveness and efficacy. In cases of recurring situations, strategies may be defined or redefined to address such situations. Participant P4 said:\textcolor{violet}{\textit{
``What we do is, in the following retrospectives, we review the improvements discussed in past retrospectives, we review those improvements and see if they are still having an impact and if we are managing to evolve in some way.''}}

Persistent situations are addressed in the current retrospective for evaluation and to define a new approach in the action plan. Participant P2 emphasized the recording of actions on the board and the recurrence of some of the points discussed in the retrospectives: \textcolor{violet}{
\textit{``We place an action item card on the board, [...] In the retrospective meeting, the first action is: Do these action items still make sense? [...] Has the action item we mapped last week been improved? [...] Is it worth keeping this action item? [...] When an action item is more critical, it usually lasts 4 or 5 sprints; [...] a smaller, more flexible action item usually lasts one, at most two sprints.''}}

On the other hand, there were reports that some teams do not monitor and routinely ignore the points from the previous Retrospective when starting the current Retrospective. Participant P5 explains: \textcolor{violet}{\textit{``We create an action, but there is no follow-up. [...] we do not revisit the one we created today to see if it is actually being followed up on.''}}
Participant P10 highlighted the lack of perception regarding monitoring and attributed it to the project manager, saying: \textcolor{violet}{\textit{``I do not do this follow-up; it is the project manager who does. But [...] what I perceive is that there is no follow-up. [...] I do not notice any monitoring. [...] this is not done by the team, nor by the manager.''}}

Some teams may use ongoing monitoring of actions during the iteration, assigning responsible members for this task, as indicated by participant P14, who stated:\textcolor{violet}{\textit{``We are using a strategy of, weekly, going through the points. At least once a week, once or twice, go through the retrospective points and check if the person is doing it.''}} Therefore, it can be observed that the retrospective meeting can provide some level of monitoring of situations and actions during the meeting itself, which frequently occurs every two weeks.

\subsection{(RQ4) What are The Benefits of Retrospective Meetings in Preventing Community Smells?}

Highlighting positive aspects can generate positive emotions in team members, producing well-being and positive reinforcement for their behaviour, which can help avoid the causes or minimize the effects of Community Smells. However, agile teams are prone to focusing on negative aspects in Retrospective Meetings, which is why it is necessary to use some dynamics to conduct the meeting~\cite{matthies2020}.

Positive aspects can be linked to critical factors for team effectiveness, which, as discussed in \cite{caballero2023}, are associated with both the causes and effects of Community Smells. In this context, participants were asked: \textit{``What has been highlighted as positive in your team's retrospective meetings?''}, considering that such points can positively contribute to the reduction of social debt.

\begin{figure*}[ht]
    \centering
    \includegraphics[width=0.9\linewidth]{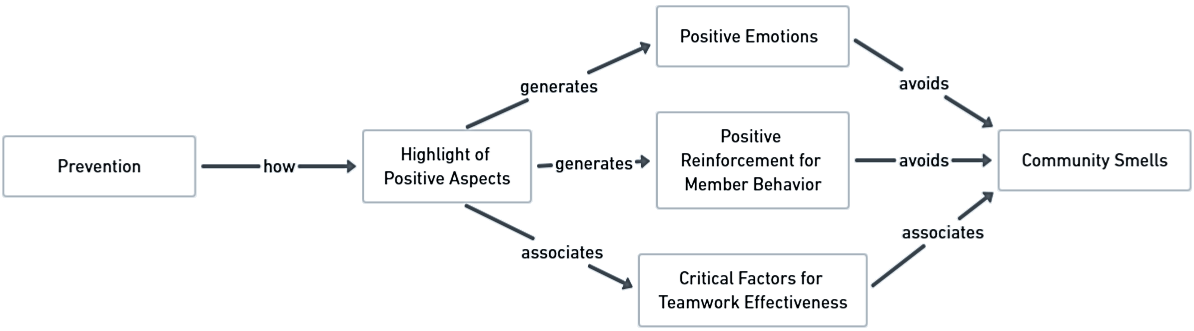}
    \caption{Prevention of Community Smells in Retrospective Meetings}
    \label{fig:prevention-csmells}
\end{figure*}

Table~\ref{tab:positivos} shows the positive aspects highlighted by the teams of the participants in this research. Among the positive points highlighted, the aspects were related to team engagement (unity, commitment, collaboration), good communication (availability, understanding, and languages in global contexts), addition of new team members, recognition of members due to performance in the iteration, expressions of gratitude among members, and the aspect of the environment being psychologically safe.
The hypothesis highlighted is that, besides being associated with the critical factors of work in the team and the causes and effects of Community Smells, the positive aspects can generate positive emotions and positive reinforcements for members, contributing to avoiding Community Smells situations. The Fig.~\ref{fig:prevention-csmells} presents these concepts and their relationships to the prevention of Community Smells.

\begin{table*}
    \centering
    \caption{Positive Aspects Highlighted in Retrospective Meetings}
    \resizebox{\textwidth}{!}{
    \begin{tabular}{|c|l|l|l|}
    \hline
        \textbf{ID} & \textbf{Positive Aspect} & \textbf{Example Segments} & \textbf{Interviews (Participant ID)}\\ \hline \hline
        % 1 & Productivity & 
        % \textit{\begin{tabular}{b}
        %      (P2) ``amount of deliveries we made''\\
        %      (P8) ``When the development team manages to deliver\\ the development early, we always congratulate them for it.''\end{tabular}} & P2, P3, P5, P6, P8, P11, P13\\ \hline
        1 & Team Engagement & 
        \textit{\begin{tabular}{l}
             (P3) ``behavior of everyone in the squad, unity, commitment''\end{tabular}} & P3, P4, P5, P6, P7, P10, P14\\ \hline
        2 & Good Communication & 
        \textit{\begin{tabular}{l}
             (P9) ``the communication part was very smooth... especially because\\ we weren't native speakers speaking to %were brazilians speaking english with
             the foreign team.'' \\
             (P14) ``people still often mention that when the\\ communication is good between the teams'' \end{tabular}} & P5, P6, P8, P9, P10, P14, P15 \\ \hline
        % 4 & Stage Completion &
        % \textit{\begin{tabular}{b}
        %      (P8) ``qwhen we manage to deliver a specific demand,\\ because we have those goals to meet.''\end{tabular}} &  P4, P8, P13, P14, P15\\ \hline
        3 & New Members & 
        \textit{\begin{tabular}{l}
             (P7) ``When a new member joins, it's a positive point\\ because it's an additional workforce.''\end{tabular}} & P4, P7, P14\\ \hline

        4 & Recognition of Members & \textit{\begin{tabular}{l}
             (P12) ``We really liked to bring personal recognition\\ to retrospectives. For example, we saw that the PO\\ worked hard in the last sprint, went after things,\\ and really supported the issue well.'' \end{tabular}} & P12, P13, P15\\ \hline
             
        % 7 & Reduction of Technical Debt & 
        % \textit{\begin{tabular}{b}
        %      ``fixing small issues that could\\ lead to bigger problems or bugs in the future.''\end{tabular}} & P2 \\ \hline
        
        5 & Gratitude & 
        \textit{\begin{tabular}{l}
             ``thanking other team members who helped you during the week\\ or the sprint''\end{tabular}} & P4\\ \hline
        6 & Psychologically Safe Environment & 
        \textit{\begin{tabular}{l}
             ``The team has a lot of trust among its members.\\ We don't seek a blame-oriented approach.\\ We had a problem. Who was responsible for it?\\ We don't adopt that. We focus on solving the problem.'' \end{tabular}} & P6\\ \hline
        % 10 & Greater Test Coverage & 
        % \textit{\begin{tabular}{b}
        %      ``The stories weren't that large, so we could have\\ more specific tests, unit tests, things like that.\\ Therefore, we had a very low bug rate.\\ This didn't delay the sprints as much and caused fewer issues\\ with what we were developing.'' \end{tabular}} & P9\\ \hline
        % 11 & Recognition of Adopted Practices & \textit{\begin{tabular}{b} ``We started leaving some test cases for them\\ and later they ended up recognizing and gaining that autonomy.\\ So the team recognized this and brought it up in the retrospective.''\end{tabular}} & P12 \\ \hline
    \end{tabular}
    }
    \label{tab:positivos}
\end{table*}

\section{Discussion}
\label{discussion}

This section presents a discussion of the results obtained, the hypotheses, and the implications of this study.

\subsection{Discussion of the results}

In the interpretative framework presented by \cite{caballero2023} to explain the origin and evolution of Community Smells, the initial stages involve experiences whose effects can be felt by members individually and by the team. Thus, such aspects can be highlighted in the Retrospective Meeting by one of the team members and discussed by all to define an improvement action.

%\textcolor{red}{The studies \cite{dingsoyr2018, martini2019} also highlight the detection of social debt by teams in retrospective meetings. However, they do not address Community Smells and do not discuss the defined actions, especially the strategies for refactoring the social debt situations addressed in this research.}

Cases of Lone Wolf, being related to specific members, are reported in a generic manner in retrospectives, attributing them to the entire team. When considering the fact that the cards (points mentioned by the members) are shared anonymously, members seem to be more motivated to share their contributions without fear. Participant P15 highlighted: \textcolor{violet}{\textit{``If it's remote and anonymous, I think people are a bit more courageous in discussing some topics, in saying some things, especially if they are not identified.''}}

The result supports the case study conducted by \cite{khanna2022}, which aimed to investigate whether online retrospective meetings are psychologically safe. It is highlighted that when team members make their contributions anonymously, the team feels freedom, trust and is more motivated to express themselves, feeling more motivated to contribute in the meeting.

In some LW cases, improvement actions are often not established by the team during the retrospective meeting but are addressed by independent initiatives of some members (in general, leaders who carry out mentoring or task coordination). It is observed that one of the justifications for not defining actions is how the situation is presented—gently, as a general alert/request to the group rather than directing it at the specific member. Another factor is that, because this Community Smell pertains to specific individuals, cases are also discussed at other moments, such as in one-on-one meetings, to avoid putting members in uncomfortable situations. Participant P12 expresses:\textcolor{violet}{\textit{
``We never pointed fingers at individuals. Since there were other meetings, like one-on-ones, the leader would communicate [...]. [...] people never mentioned the person's name directly.''}}

Situations of Radio Silence that involve stakeholders and other teams are generally more difficult to resolve, making them recurrent in retrospective meetings. %This may explain why it is one of the most frequently reported Community Smells by participants in this research.
The case study conducted by \cite{Pikkarainen2008} investigated the impact of agile practices on communication within software development teams, internally between developers and project leaders, and in the interface between the development team and stakeholders (clients, testers, other development teams). Among the findings, it is noted that in contexts with multiple stakeholders external to the team, the lack of an adequate communication mechanism can sometimes hinder communication.

The communication difficulties indicated by \cite{Pikkarainen2008} can be exemplified by the Radio Silence reports obtained in this research. For example, participant P6 describes the difficulty in communication and resolution of a situation involving stakeholders, where different methods were attempted but were not effective in obtaining information related to a functional requirement. Participant P6 said:  \textcolor{violet}{
\textit{``We had some requirements, but they were not very well documented [...] we tried to communicate. They asked us to exchange emails, but the information still wasn’t clear. We even had to undo the implementation because [...] was not exactly what the client wanted.''}}

Regarding the strategies defined for Organizational Silo and Black Cloud, most of them are present in other studies, such as \cite{caballero2023, catolino2020, tahsin2022, tamburri2013}, which demonstrate their widespread use and effectiveness for these situations.

Retrospective meetings are used by teams to monitor specific problem situations. Teams may use the time to assess the actions defined and check if the situations still occur or, simply, focus on events from the past iteration without revisiting actions defined in the previous meeting.

Both scenarios—retrospectives with review and retrospectives without review of the previous retrospective items—seem to be equivalent in monitoring Community Smells when considering the properties of this type of smell, such as harmlessness and latency. That is, the effects take some time to manifest and are temporarily harmless until they become more frequent or prevalent \cite{caballero2023, palomba2021, tamburri2019b, tamburri2016, tamburri2015}. Thus, retrospective moments can emphasize recurring situations and ineffective actions discussed in the ongoing retrospective, with the aim of establishing new actions or adjusting them for the situations perceived by the team members during the closing iteration.

Specific monitoring of actions during the iteration may not occur or may not be visible to all team members. Leaders or managers may independently monitor situations and actions, or teams may assign individuals responsible for executing and monitoring certain actions during the next iteration.

In the retrospective meeting, the moment when the team lists the positive points, besides helping the team not to focus solely on the negatives~\cite{matthies2019}, the emphasis on positive points can serve as an indicator of social factors within the team and act as a reinforcing element for the appropriate behaviour of members.
In \cite{caballero2023}, a set of critical factors for team effectiveness \cite{bell2018}\cite{salas2015} was used, relating them to the causes of Community Smells, as well as highlighting the influence flow of the effects of Community Smells on these factors\footnote{A definition of each critical factor can be found in Table 1 in \cite[p. 3]{caballero2023}}.

\begin{figure*}[ht]
    \centering
    \includegraphics[width=1.0\linewidth]{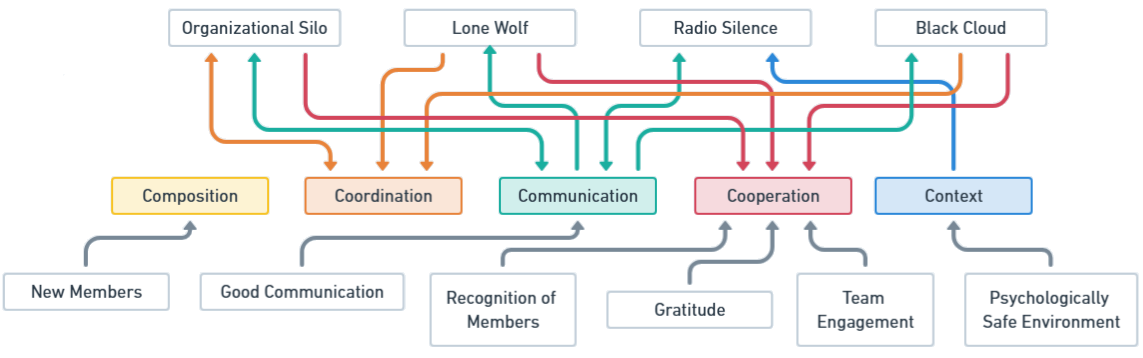}
    \caption{Association of the Positive Aspects and Community Smells through Critical Factors for Teamwork Effectiveness.}
    \label{fig:aspectos-positivos}
\end{figure*}

Similarly, using the critical factors as codes, associating the identified positive aspects with the critical factors for team effectiveness, we can observe that the positive aspects highlighted in the retrospective meetings align with the most frequent factors causing Community Smells according to \cite[p. 15]{caballero2023} — Context, Communication, Composition, Coordination, and Cooperation. Therefore, highlighting these aspects as positive in the retrospective meetings can indicate health (absence of causes and/or effects) concerning the circumstances of Community Smells.
The Fig.~\ref{fig:aspectos-positivos} illustrates the association between the positive aspects identified in the research and the critical factors for team effectiveness -- which are associated with the causes and effects of the Community Smells addressed in this research. The critical factor ``Composition'' was not associated with the causes or effects of the four Community Smells approached in this study.

The study \cite{madampe2024} argues that practitioners experience various negative emotions when working on software projects, and these emotions can interfere with individual productivity and impact team productivity. In \cite{carneiro2020}, it is shown that the feelings of developers in software projects impact practices and artefacts, and that positive feelings tend to positively impact practices, e.g., increasing engagement and shortening problem-solving time; while negative feelings negatively affect practices but can also have positive effects, e.g., reducing downtime or making the review process more thorough.

In another empirical study \cite{graziotin2018}, the internal consequences for software developers and external consequences of happiness and unhappiness are observed to understand the impact on software engineering activities. The results show, among other aspects, that happiness leads to increased performance, motivation, creativity, self-fulfilment, and the perception of a positive atmosphere by software developers, with external consequences linked to productivity, sustainable flow, increased collaboration, and code quality.

The Retrospective Meeting enables the detection of Community Smells through the relevance given, by at least one of the members, to social debt situations, bringing them to the meeting. In \cite{tahsin2022}, among other factors, the perception of the practitioner in the proposed metric to measure the readiness of team for refactoring Community Smells (\textit{Refactoring Readiness}) is considered. In this context, it is believed that such awareness among members can highlight social debt situations in retrospective meetings, enhancing the detection and definition of strategies for refactoring.

Regarding the definition of refactoring strategies, it is noteworthy that in retrospectives, teams focus on problematic situations rather than necessarily on the strategies defined as action plans. In this sense, it can only be stated that the Community Smells highlighted in the teams' retrospectives were generally addressed. Furthermore, the resolution of these situations cannot be solely attributed to the strategies defined by the teams, as there is no data on the implementation of these actions. However, many of the strategies defined are present in the literature, such as mentoring, team restructuring, and the creation or adjustment of communication plans \cite{catolino2020, tamburri2013, tahsin2022}, which have proven effectiveness for the Community Smells addressed in research by \cite{caballero2023}.

Since teams conduct retrospective meetings at the end of each iteration (approximately every two weeks), and situations that characterize Community Smells are not constant or visible at all times, detecting and monitoring the team's state concerning social debt can be more efficient by anticipating detection and mitigation. This efficiency contrasts with the 3-month time window used in other studies exploring the occurrence of Community Smells \cite{magnoni2016, palomba2021, tamburri2021, huang2022a}.

Given the relationship between positive aspects listed in retrospective meetings (highlighted in this study), the factors of effectiveness of teamwork and the causes and effects of Community Smells \cite{caballero2023}, considering further the impacts of positive emotions on the productivity of individuals and their teams \cite{graziotin2018, carneiro2020, madampe2024}, it is pointed out that the positive aspects can generate positive emotions in the members involved and impact productivity, as well as provide positive reinforcement for behaviours that contribute to avoiding social debt and circumstances of Community Smells.

\subsection{Implications}
\label{implications}

The results found have some implications for both researchers and practitioners in the field of software engineering.

\textit{For practitioners:} The study revealed how retrospective meetings show Community Smells in agile development teams, highlighting the importance of such meetings for continuous improvement of teams in different aspects, including social aspects. However, barriers such as low psychological safety (presence of blame, shame, non-anonymous) can inhibit member participation and reduce contributions in the identification, definition and monitoring of Community Smells refactoring strategies. In this sense, teams should cultivate trust and respect among members to strengthen collaboration, increase the quality of the retrospective meeting and refactoring Community Smells.

Refactoring and monitoring strategies are not always defined or carried out. Thus, the teams must act systematically, ensuring that all situations highlighted in the retrospective meeting have a defined action and are registered in the team board. They must also perform the review of such items in the following retrospective. The highlighting of positive aspects should not be neglected, as they can contribute to the prevention of Community Smells, acting as reinforcements for desired behaviours and generating positive emotions in members.

\textit{For researchers:} This work was limited to 4 of 30 Community Smells cataloged by Caballero-Espinosa, Carver, and Stowers\cite{caballero2023}. Thus, it does not address strategies defined for other Community Smells, among them, Community Smells without strategies present in the literature, with strategies without proven or partially proven effectiveness. Researchers can act in the reproduction of this study using other smells or replicate with practitioners from other nationalities -- in this study, all practitioners are Brazilians working in teams formed by only Brazilians, except for one, who is working in a European team. New research may investigate the effectiveness of strategies defined in retrospective meetings, as well as deepen the insights on the impact of positive aspects in preventing Community Smells.

%\section{Threats to validity}
\section{Trustworthiness of the Study}
\label{validade}

For validation of findings from qualitative research, we adhere to Creswell's~\cite{creswell2014} guidance by considering the credibility, qualitative reliability, and transferability of the findings.

\textbf{Credibility.} Our study involved 15 participants, which may characterize the risk of not being truly representative of the technology industry and teams. We have tried to minimize this risk by selecting a set of diverse participants from a variety of companies and backgrounds. However, we felt saturation was reached in coding 12\textsuperscript{th} interviews, as we did not determine any new codes or themes when analyzing the last four interviews -- consistent with \cite{guest2006} and \cite{thirycherques2009}. The use of LinkedIn to recruit participants allows access to different profiles of practitioners, however, the connection with the profile and the exchange of messages is asynchronous, requiring waiting time for actions of the recruited, in addition to refusal or inaction by the recruited user. Due to these limitations, some participants were recruited by indications of other researchers.

\textbf{Reliability.} We have documented the study design in Section~\ref{method}. Two researchers were involved in the data analysis and through regular meetings used a consensus mechanism to resolve any disagreements in the coding of segments referring to Community Smells. Furthermore, a third researcher was involved across the entire analysis to monitor the process being followed and verified the analysis and its results.

Although the results highlight the personal perception of members about community smells, the format (simplicity, flexibility, and frequency) and the intention (continuous improvement) of the retrospective meeting favour the exposure and decision-making regarding the aspects considered relevant by at least one of the team members. New studies with data from observations and records of team retrospective meetings can further highlight the relevance of the personal perception of members of the team about social situations and highlight the trade-offs of the retrospective meeting on the management of social debt.

The structure and conducting of retrospective meetings experienced by participants can influence results because, in the structure, the topic ``what could be improved'' can involve different aspects; sometimes, teams focus on technical issues and not address issues related to social factors. Likewise, anonymity and psychological safety can provide members with more confidence in exposing situations related to social debt.

\textbf{Transferability.} Our participants worked in several different teams, based in single country, so our results are likely transferable to software teams working in this country. However, work practices and norms of the country likely have some impact on the experiences participants described to us. The study explored only four of the 30 Community Smells cataloged. Thus, it was limited to the perception of single nationality practitioners in their teams regarding the reduced number of Community Smells, and it is not possible to state whether such circumstances are highlighted in retrospective meetings by practitioners with other nationalities or if the results apply to other Community Smells.

\section{Conclusions}
\label{conclusion}

The Retrospective Meeting enables the identification of Community Smells, the definition and monitoring of refactoring strategies, and can also contribute to prevention through the emphasis on positive aspects.
As it is constituted by points raised by team members, the Retrospective Meeting can be influenced by the individual perceptions of the members, the criticality of the effects on individual and/or team work, and the context or culture to which the team is subjected. In some teams, due to these factors, retrospective meetings may focus more on technical and process aspects.
Regarding refactoring strategies and potential preventive practices, further investigations are needed to understand how such actions impact the situations and what the trade-offs are for the team, project, and organization.

Future research could explore the role of Retrospective Meetings in specific operational contexts, such as DevOps teams, as well as in large-scale agile methodologies contexts, to highlight the detection of Community Smells and the respective refactoring strategies adopted.
Finally, this research contributes to the understanding of addressing social debt issues in software engineering, specifically Community Smells in agile teams, providing practical insights for retrospective meetings and a foundation for future research on the topic.

\bibliographystyle{IEEEtran}
% (plain),unsrt,alpha,abbrv.
\bibliography{conference_101719}

\begin{thebibliography}{10}
\providecommand{\url}[1]{#1}
\csname url@samestyle\endcsname
\providecommand{\newblock}{\relax}
\providecommand{\bibinfo}[2]{#2}
\providecommand{\BIBentrySTDinterwordspacing}{\spaceskip=0pt\relax}
\providecommand{\BIBentryALTinterwordstretchfactor}{4}
\providecommand{\BIBentryALTinterwordspacing}{\spaceskip=\fontdimen2\font plus
\BIBentryALTinterwordstretchfactor\fontdimen3\font minus \fontdimen4\font\relax}
\providecommand{\BIBforeignlanguage}[2]{{%
\expandafter\ifx\csname l@#1\endcsname\relax
\typeout{** WARNING: IEEEtran.bst: No hyphenation pattern has been}%
\typeout{** loaded for the language `#1'. Using the pattern for}%
\typeout{** the default language instead.}%
\else
\language=\csname l@#1\endcsname
\fi
#2}}
\providecommand{\BIBdecl}{\relax}
\BIBdecl

\bibitem{tamburri2013}
D.~A. Tamburri, P.~Kruchten, P.~Lago, and H.~van Vliet, ``What is social debt in software engineering?'' in \emph{2013 6th International Workshop on Cooperative and Human Aspects of Software Engineering (CHASE)}, May 2013, pp. 93--96.

\bibitem{catolino2020}
\BIBentryALTinterwordspacing
G.~Catolino, F.~Palomba, D.~A. Tamburri, A.~Serebrenik, and F.~Ferrucci, ``\BIBforeignlanguage{English}{Refactoring community smells in the wild: The practitioner's field manual},'' in \emph{\BIBforeignlanguage{English}{Proceedings of the ACM/IEEE 42nd International Conference on Software Engineering: Software Engineering in Society}}, ser. ICSE-SEIS '20.\hskip 1em plus 0.5em minus 0.4em\relax New York, NY, USA: Association for Computing Machinery, Oct 2020, Conference paper, p. 25–34, cited by: 7. [Online]. Available: \url{https://doi.org/10.1145/3377815.3381380}
\BIBentrySTDinterwordspacing

\bibitem{tamburri2015}
\BIBentryALTinterwordspacing
D.~A. Tamburri, P.~Kruchten, P.~Lago, and H.~v. Vliet, ``Social debt in software engineering: insights from industry,'' \emph{Journal of Internet Services and Applications}, vol.~6, no.~1, p.~10, 2015. [Online]. Available: \url{https://doi.org/10.1186/s13174-015-0024-6}
\BIBentrySTDinterwordspacing

\bibitem{tamburri2021}
\BIBentryALTinterwordspacing
D.~A. Tamburri, F.~Palomba, and R.~Kazman, ``\BIBforeignlanguage{English}{Exploring community smells in open-source: An automated approach},'' \emph{\BIBforeignlanguage{English}{IEEE Transactions on Software Engineering}}, vol.~47, no.~3, pp. 630--652, March 2021, cited by: 20; All Open Access, Green Open Access. [Online]. Available: \url{https://www.scopus.com/inward/record.uri?eid=2-s2.0-85062150097&doi=10.1109%2fTSE.2019.2901490&partnerID=40&md5=f9d049afe174884268c53fc7aceb0b06}
\BIBentrySTDinterwordspacing

\bibitem{palomba2021b}
F.~Palomba, D.~Andrew~Tamburri, F.~Arcelli~Fontana, R.~Oliveto, A.~Zaidman, and A.~Serebrenik, ``Beyond technical aspects: How do community smells influence the intensity of code smells?'' \emph{IEEE Transactions on Software Engineering}, vol.~47, no.~1, pp. 108--129, 2021.

\bibitem{caballero2023}
\BIBentryALTinterwordspacing
E.~Caballero-Espinosa, J.~C. Carver, and K.~Stowers, ``Community smells—the sources of social debt: A systematic literature review,'' \emph{Information and Software Technology}, vol. 153, p. 107078, 2023. [Online]. Available: \url{https://www.sciencedirect.com/science/article/pii/S0950584922001872}
\BIBentrySTDinterwordspacing

\bibitem{agilepraticeguide2017}
A.~Alliance, \emph{Agile Practice Guide, Project Management Institute, 2017: Agile Practice Guide}.\hskip 1em plus 0.5em minus 0.4em\relax Bukupedia, 2017, vol.~1.

\bibitem{agiletrenches2015}
H.~Kniberg, \emph{Scrum and XP from the Trenches}.\hskip 1em plus 0.5em minus 0.4em\relax Lulu. com, 2015.

\bibitem{stateagile2016}
V.~One, ``11th state of agile report,'' Version One, Tech. Rep., 2016.

\bibitem{scrumguide2020}
\BIBentryALTinterwordspacing
K.~Schwaber and J.~Sutherland, ``Scrum guide,'' Scrumguides.org, 11 2020. [Online]. Available: \url{https://scrumguides.org/scrum-guide.html}
\BIBentrySTDinterwordspacing

\bibitem{kua2012retrohandbook}
P.~Kua, \emph{The Retrospective Handbook}.\hskip 1em plus 0.5em minus 0.4em\relax Leanpub, 2013.

\bibitem{derby2006agileretrospectives}
E.~Derby, D.~Larsen, and K.~Schwaber, \emph{Agile retrospectives: Making good teams great}.\hskip 1em plus 0.5em minus 0.4em\relax Pragmatic Bookshelf, 2006.

\bibitem{caroli2020funretrospectives}
P.~Caroli and T.~C. Coimbra, \emph{FunRetrospectives: activities and Ideas for making agile retrospectives more engaging}.\hskip 1em plus 0.5em minus 0.4em\relax Editora Caroli, 2020.

\bibitem{martini2019}
A.~Martini, V.~Stray, and N.~B. Moe, ``Technical-, social- and process debt in large-scale agile: An exploratory case-study,'' in \emph{Agile Processes in Software Engineering and Extreme Programming -- Workshops}, R.~Hoda, Ed.\hskip 1em plus 0.5em minus 0.4em\relax Cham: Springer International Publishing, 2019, pp. 112--119.

\bibitem{scrumguide2022bok}
SCRUMStudy, \emph{A Guide to the Scrum Body of Knowledge (SBOK® Guide) – Fourth edition}.\hskip 1em plus 0.5em minus 0.4em\relax Scrumstudy, A Brand Of Vmedu, Inc, 2022.

\bibitem{Tamburri2019a}
\BIBentryALTinterwordspacing
D.~A. Tamburri, R.~Kazman, and W.-J. van~den Heuvel, ``Splicing community and software architecture smells in agile teams: An industrial study,'' in \emph{Hawaii International Conference on System Sciences}, 2019. [Online]. Available: \url{https://api.semanticscholar.org/CorpusID:102351977}
\BIBentrySTDinterwordspacing

\bibitem{charmaz2009}
\BIBentryALTinterwordspacing
K.~Charmaz, \emph{A constru{\c{c}}{\~a}o da teoria fundamentada: Guia Pr{\'a}tico para An{\'a}lise Qualitativa}, ser. M{\'e}todos de Pesquisa.\hskip 1em plus 0.5em minus 0.4em\relax Penso, 2009. [Online]. Available: \url{https://books.google.com.br/books?id=offC0wDYzC4C}
\BIBentrySTDinterwordspacing

\bibitem{sarmento2022}
\BIBentryALTinterwordspacing
C.~Sarmento, T.~Massoni, A.~Serebrenik, G.~Catolino, D.~Tamburri, and F.~Palomba, ``\BIBforeignlanguage{English}{Gender diversity and community smells: A double-replication study on brazilian software teams},'' in \emph{\BIBforeignlanguage{English}{2022 IEEE International Conference on Software Analysis, Evolution and Reengineering (SANER)}}.\hskip 1em plus 0.5em minus 0.4em\relax Institute of Electrical and Electronics Engineers Inc., March 2022, Conference paper, pp. 273--283, cited by: 2. [Online]. Available: \url{https://www.scopus.com/inward/record.uri?eid=2-s2.0-85133692894&doi=10.1109%2fSANER53432.2022.00043&partnerID=40&md5=e7fabf15f8c5c0dc44830467eb9dd6fd}
\BIBentrySTDinterwordspacing

\bibitem{tahsin2022}
\BIBentryALTinterwordspacing
N.~Tahsin and K.~Sakib, ``\BIBforeignlanguage{English}{Refactoring community smells: An empirical study on the software practitioners of bangladesh},'' in \emph{\BIBforeignlanguage{English}{2022 29th Asia-Pacific Software Engineering Conference (APSEC)}}, vol. 2022-December.\hskip 1em plus 0.5em minus 0.4em\relax IEEE Computer Society, Dec 2022, Conference paper, pp. 422--426, cited by: 0. [Online]. Available: \url{https://www.scopus.com/inward/record.uri?eid=2-s2.0-85149172008&doi=10.1109%2fAPSEC57359.2022.00055&partnerID=40&md5=21eb50e80022f084d7eb0dac443ad1cf}
\BIBentrySTDinterwordspacing

\bibitem{law2005}
\BIBentryALTinterwordspacing
A.~Law and R.~Charron, ``Effects of agile practices on social factors,'' in \emph{Proceedings of the 2005 Workshop on Human and Social Factors of Software Engineering}, ser. HSSE '05.\hskip 1em plus 0.5em minus 0.4em\relax New York, NY, USA: Association for Computing Machinery, 2005, p. 1–5. [Online]. Available: \url{https://doi.org/10.1145/1083106.1083115}
\BIBentrySTDinterwordspacing

\bibitem{mchugh2012}
O.~McHugh, K.~Conboy, and M.~Lang, ``Agile practices: The impact on trust in software project teams,'' \emph{IEEE Software}, vol.~29, no.~3, pp. 71--76, 2012.

\bibitem{muller2021}
D.~Müller, M.~Kropp, C.~Anslow, and A.~Meier, ``The effects on social support and work engagement with scrum events,'' in \emph{2021 IEEE/ACM 13th International Workshop on Cooperative and Human Aspects of Software Engineering (CHASE)}, 2021, pp. 101--104.

\bibitem{dingsoyr2018}
T.~Dings{\o}yr, M.~Mikalsen, A.~Solem, and K.~Vestues, ``Learning in the large - an exploratory study of retrospectives in large-scale agile development,'' in \emph{Agile Processes in Software Engineering and Extreme Programming}, J.~Garbajosa, X.~Wang, and A.~Aguiar, Eds.\hskip 1em plus 0.5em minus 0.4em\relax Cham: Springer International Publishing, 2018, pp. 191--198.

\bibitem{creswell2014}
\BIBentryALTinterwordspacing
J.~Creswell, \emph{Investiga{\c{c}}{\~a}o Qualitativa e Projeto de Pesquisa - 3.ed.: Escolhendo entre Cinco Abordagens}, ser. M{\'e}todos de Pesquisa.\hskip 1em plus 0.5em minus 0.4em\relax Penso Editora, 2014. [Online]. Available: \url{https://books.google.com.br/books?id=Ymi5AwAAQBAJ}
\BIBentrySTDinterwordspacing

\bibitem{sampieri2013}
\BIBentryALTinterwordspacing
R.~H. Sampieri, C.~F. Collado, and M.~D. P.~B. Lucio, \emph{Metodologia de Pesquisa}, ser. M{\'e}todos de Pesquisa.\hskip 1em plus 0.5em minus 0.4em\relax AMGH Editora, 2013. [Online]. Available: \url{https://books.google.com.br/books?id=AKU5AgAAQBAJ}
\BIBentrySTDinterwordspacing

\bibitem{matthies2020}
\BIBentryALTinterwordspacing
C.~Matthies and F.~Dobrigkeit, ``Towards empirically validated remedies for scrum retrospective headaches,'' in \emph{Proceedings of the 53rd Hawaii International Conference on System Sciences}, ser. HICSS.\hskip 1em plus 0.5em minus 0.4em\relax Hawaii International Conference on System Sciences, 2020. [Online]. Available: \url{a://dx.doi.org/10.24251/HICSS.2020.762}
\BIBentrySTDinterwordspacing

\bibitem{khanna2022}
D.~Khanna and X.~Wang, ``Are your online agile retrospectives psychologically safe? the usage of online tools,'' in \emph{Agile Processes in Software Engineering and Extreme Programming}, V.~Stray, K.-J. Stol, M.~Paasivaara, and P.~Kruchten, Eds.\hskip 1em plus 0.5em minus 0.4em\relax Cham: Springer International Publishing, 2022, pp. 35--51.

\bibitem{Pikkarainen2008}
\BIBentryALTinterwordspacing
M.~Pikkarainen, J.~Haikara, O.~Salo, P.~Abrahamsson, and J.~Still, ``The impact of agile practices on communication in software development,'' \emph{Empirical Softw. Engg.}, vol.~13, no.~3, p. 303–337, jun 2008. [Online]. Available: \url{https://doi.org/10.1007/s10664-008-9065-9}
\BIBentrySTDinterwordspacing

\bibitem{palomba2021}
\BIBentryALTinterwordspacing
F.~Palomba and D.~A. Tamburri, ``Predicting the emergence of community smells using socio-technical metrics: A machine-learning approach,'' \emph{Journal of Systems and Software}, vol. 171, p. 110847, 2021. [Online]. Available: \url{https://www.sciencedirect.com/science/article/pii/S0164121220302375}
\BIBentrySTDinterwordspacing

\bibitem{tamburri2019b}
\BIBentryALTinterwordspacing
D.~A. Tamburri, ``\BIBforeignlanguage{English}{Software architecture social debt: Managing the incommunicability factor},'' \emph{\BIBforeignlanguage{English}{IEEE Transactions on Computational Social Systems}}, vol.~6, no.~1, pp. 20--37, Feb 2019, cited by: 14; All Open Access, Green Open Access. [Online]. Available: \url{https://www.scopus.com/inward/record.uri?eid=2-s2.0-85061655975&doi=10.1109%2fTCSS.2018.2886433&partnerID=40&md5=e6693e173e86e6f5ee46cff0ae2f0a06}
\BIBentrySTDinterwordspacing

\bibitem{tamburri2016}
\BIBentryALTinterwordspacing
D.~A. Tamburri, R.~Kazman, and H.~Fahimi, ``\BIBforeignlanguage{English}{The architect's role in community shepherding},'' \emph{\BIBforeignlanguage{English}{IEEE Software}}, vol.~33, no.~6, p. 70 – 79, 2016, cited by: 49. [Online]. Available: \url{https://www.scopus.com/inward/record.uri?eid=2-s2.0-84994403514&doi=10.1109%2fMS.2016.144&partnerID=40&md5=903721c96045a6521a1d30f38a44c46e}
\BIBentrySTDinterwordspacing

\bibitem{matthies2019}
C.~Matthies, F.~Dobrigkeit, and A.~Ernst, ``Counteracting agile retrospective problems with retrospective activities,'' in \emph{Systems, Software and Services Process Improvement}, A.~Walker, R.~V. O'Connor, and R.~Messnarz, Eds.\hskip 1em plus 0.5em minus 0.4em\relax Cham: Springer International Publishing, 2019, pp. 532--545.

\bibitem{bell2018}
S.~T. Bell, S.~G. Brown, A.~Colaneri, and N.~Outland, ``Team composition and the abcs of teamwork.'' \emph{American Psychologist}, vol.~73, pp. 349--362, 05 2018.

\bibitem{salas2015}
E.~Salas, M.~L. Shuffler, A.~L. Thayer, W.~L. Bedwell, and E.~H. Lazzara, ``Understanding and improving teamwork in organizations: A scientifically based practical guide,'' \emph{Human Resource Management}, vol.~54, pp. 599--622, 10 2015.

\bibitem{madampe2024}
K.~Madampe, R.~Hoda, and J.~Grundy, ``Addressing bad feelings in agile software project contexts,'' \emph{IEEE Software}, no.~01, pp. 1--6, mar 2024.

\bibitem{carneiro2020}
G.~de~Figueiredo~Carneiro and R.~C. J{\'u}nior, ``Investigating the impact of developers sentiments on software projects,'' in \emph{17th International Conference on Information Technology--New Generations (ITNG 2020)}, S.~Latifi, Ed.\hskip 1em plus 0.5em minus 0.4em\relax Cham: Springer International Publishing, 2020, pp. 257--263.

\bibitem{graziotin2018}
\BIBentryALTinterwordspacing
D.~Graziotin, F.~Fagerholm, X.~Wang, and P.~Abrahamsson, ``What happens when software developers are (un)happy,'' \emph{Journal of Systems and Software}, vol. 140, pp. 32--47, 2018. [Online]. Available: \url{https://www.sciencedirect.com/science/article/pii/S0164121218300323}
\BIBentrySTDinterwordspacing

\bibitem{magnoni2016}
S.~Magnoni, ``An approach to measure community smells in software development communities,'' Ph.D. dissertation, Politecnico di Milano, 2016.

\bibitem{huang2022a}
\BIBentryALTinterwordspacing
Z.-J. Huang, Z.-Q. Shao, G.-S. Fan, H.-Q. Yu, X.-G. Yang, and K.~Yang, ``\BIBforeignlanguage{English}{Community smell occurrence prediction on multi-granularity by developer-oriented features and process metrics},'' \emph{\BIBforeignlanguage{English}{Journal of Computer Science and Technology}}, vol.~37, no.~1, p. 182 – 206, FEB 2022, cited by: 3. [Online]. Available: \url{https://www.scopus.com/inward/record.uri?eid=2-s2.0-85125323743&doi=10.1007%2fs11390-021-1596-1&partnerID=40&md5=8e4c15f6f3181100378d973e884107f4}
\BIBentrySTDinterwordspacing

\bibitem{guest2006}
\BIBentryALTinterwordspacing
G.~Guest, A.~Bunce, and L.~Johnson, ``How many interviews are enough?: An experiment with data saturation and variability,'' \emph{Field Methods}, vol.~18, no.~1, pp. 59--82, 2006. [Online]. Available: \url{https://doi.org/10.1177/1525822X05279903}
\BIBentrySTDinterwordspacing

\bibitem{thirycherques2009}
H.~R. Thirycherques, ``SaturaÇÃo em pesquisa qualitativa: Estimativa empÍrica de dimensionamento,'' \emph{Revista Brasileira de Pesquisas de Marketing, Opinião e Mídia}, vol.~2, pp. 20--27, 2009.

\end{thebibliography}
\end{document}